\documentclass[%
reprint,
superscriptaddress,
 amsmath,amssymb,
 aps,
]{revtex4-2}

\usepackage{graphicx}% Include figure files
\usepackage{dcolumn}% Align table columns on decimal point
\usepackage{bm}% bold math
\usepackage{color}
\usepackage{xcolor}

\newcommand*{\rttensortwo}[1]{\overline{\overline{#1}}}
\newcommand{\probP}{\text{I\kern-0.15em P}}
\begin{document}

\preprint{APS/123-QED}

\title{Reducing effective system  dimensionality with long-range collective dipole-dipole interactions}

% Force line breaks with \\

\author{Ashwin K. Boddeti}
\affiliation{Elmore Family School of Electrical and Computer Engineering, Purdue University,
West Lafayette, Indiana 47907, USA}
 \affiliation{Birck Nanotechnology Center, Purdue University,
West Lafayette, Indiana 47907, USA}%Lines break automatically or can be forced with \\
% \email{tboddeti@purdue.edu}
\author{Yi Wang}
\affiliation{Graduate Program in Applied Physics,
Northwestern University, Evanston, IL, 60208 USA}
\author{Xitlali G. Juarez}
\affiliation{Department of Materials Science and
Engineering, Northwestern University, Evanston, Illinois 60208,
USA}%
\author{Alexandra Boltasseva}
\affiliation{Elmore Family School of Electrical and Computer Engineering, Purdue University,
West Lafayette, Indiana 47907, USA}
 \affiliation{Birck Nanotechnology Center, Purdue University,
West Lafayette, Indiana 47907, USA}
\author{Teri W. Odom}
\affiliation{Graduate Program in Applied Physics,
Northwestern University, Evanston, IL, 60208 USA},
\affiliation{Department of Materials Science and
Engineering, Northwestern University, Evanston, Illinois 60208,
USA}
\affiliation{Department of
Chemistry, Northwestern University, Evanston, Illinois 60208, USA}
\author{Vladimir Shalaev}
\affiliation{Elmore Family School of Electrical and Computer Engineering, Purdue University,
West Lafayette, Indiana 47907, USA}
 \affiliation{Birck Nanotechnology Center, Purdue University,
West Lafayette, Indiana 47907, USA}
\author{Hadiseh Alaeian}
\affiliation{Elmore Family School of Electrical and Computer Engineering, Purdue University,
West Lafayette, Indiana 47907, USA}
 \affiliation{Birck Nanotechnology Center, Purdue University,
West Lafayette, Indiana 47907, USA}
\affiliation{School of Physics and Astronomy, Purdue University,
West Lafayette, Indiana 47907, USA}
\author{Zubin Jacob} \email{zjacob@purdue.edu \\ www.electrodynamics.org} 
\affiliation{Elmore Family School of Electrical and Computer Engineering, Purdue University,
West Lafayette, Indiana 47907, USA}  \affiliation{Birck Nanotechnology Center, Purdue University,
West Lafayette, Indiana 47907, USA}  

\date{\today}
% It is always \today, today, but any date may be explicitly specified

\begin{abstract}

Dimensionality plays a crucial role in long-range dipole-dipole interactions (DDIs). We demonstrate that a resonant nanophotonic structure modifies the apparent dimensionality in an interacting ensemble of emitters, as revealed by population decay dynamics. Our measurements on a dense ensemble of interacting quantum emitters in a resonant nanophotonic structure with long-range DDIs reveal an effective dimensionality reduction to $\bar{d} = 2.20 (12)$, despite the emitters being distributed in 3D. This contrasts the homogeneous environment, where the apparent dimension is $\bar{d} = 3.00$. Our work presents a promising avenue to manipulate dimensionality in an ensemble of interacting emitters.

\end{abstract}

\maketitle
\emph{Introduction- }In a dense ensemble of interacting emitters, each emitter perceives the other neighboring emitters via position-dependent dipole-dipole interactions (DDIs). The role of geometry in such position-dependent collective interactions between an ensemble of emitters has been of fundamental interest \cite{davis2021probing, dwyer2022probing, broholm2020quantum, yao2018quantum, chomaz2019long, sierra2022dicke, wittmann2020enhancing}. Controlling the dimensionality is appealing as a lower-dimensional emitter geometry shows strong quantum fluctuations \cite{samajdar2020complex}. This can potentially provide a host of benefits in realizing platforms to probe long-range interactions \cite{davis2021probing, dwyer2022probing}, quantum phases such as quantum spin-liquids \cite{broholm2020quantum, yao2018quantum},  transient super solid behavior \cite{chomaz2019long}, quantum phase transition in transverse Ising models \cite{schmitt2022quantum}, provide an advantage in quantum sensing applications, in mitigating decoherence \cite{davis2021probing, chomaz2019long}, and in long-range energy transport of delocalized excitons \cite{wittmann2020enhancing}. More recently, interesting physical effects on Dicke superradiance in 1D, 2D, and 3D arrays of atoms have been theoretically predicted \cite{sierra2022dicke}. Thus, realizing a lower-dimensional system supporting long-range DDIs is of significant importance.
\par 
While 1D and 2D interacting ensembles of emitters have been realized in cold-atom systems, it remains largely unexplored in solid-state platforms. Only recent efforts demonstrating a thin layer of emitters (NV - P1 centers) have paved the way for realizing lower-dimensional systems in solid states \cite{davis2021probing, dwyer2022probing}. The P1 system's many-body noise is characterized by the decoherence of NV center probe spins and shows stretched exponential decay dynamics \cite{davis2021probing}. 
 \par
As DDIs are mediated by the underlying electromagnetic fields, tailoring them provides an alternative route to manipulate the apparent dimensionality. Recently, interfacing quantum emitters with light within nanophotonic structures has provided the means to control and study collective DDIs \cite{chang2018colloquium}. This led to the demonstration of long-range resonance energy transfer in incoherent systems \cite{newman2018observation, boddeti2021long}, and sub-and super-radiant emission dynamics in coherent systems \cite{tiranov2022coherent, skljarow2022purcell, goban2015superradiance, hood2016atom}. 
\begin{figure*}
    \centering
    \includegraphics[width = 0.9\textwidth]{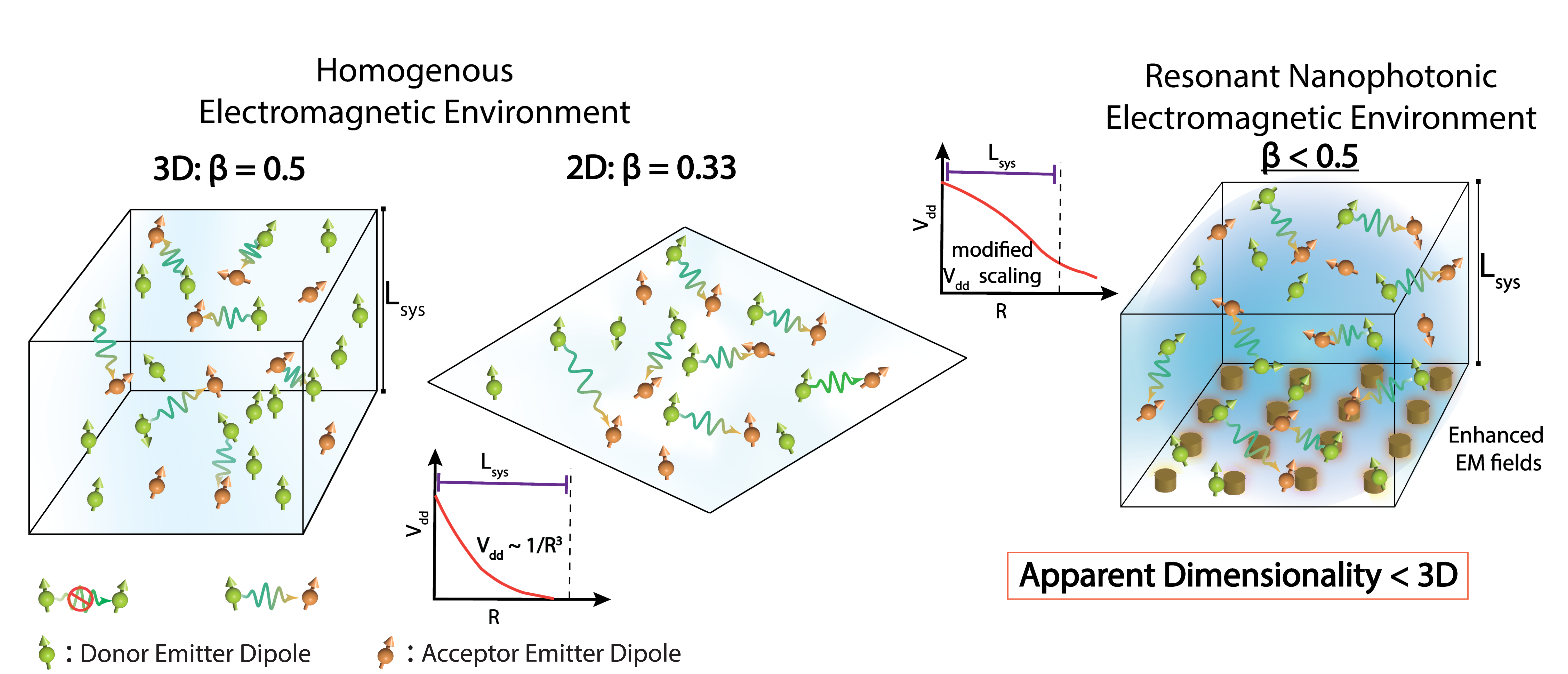}
    \caption{The illustration depicts the concept of apparent dimensionality of an interacting ensemble of emitters. The apparent dimensionality is related to the non-integer exponent of time in the fluorescence decay dynamics $I(t)/I_0 = exp(-\gamma_Dt) exp(-\alpha t^\beta)$.  In a homogeneous environment, $\beta = 0.5 (0.33)$ for the 3D 
 (2D) spatial distribution of emitters. A resonant nanophotonic environment modifies the spatial distribution of the neighboring emitters sensed by each interacting emitter which results in the modification of the temporal decay dynamics. This reduces the apparent dimensionality experienced by the interacting emitters, which is reflected in the non-integer exponent, $\beta < 0.5$, though, the emitters are distributed in a 3D volume. The green dipoles represent the donor emitters and the orange represents acceptor dipoles.}
    \label{fig1Concept}
\end{figure*}
\par
Here we modify the apparent dimensionality using a nanophotonic structure that supports dispersive delocalized resonant modes that mediate the interactions. These modes lead to modification of the spatial distribution of the perceived neighboring emitters. We experimentally probe the apparent dimensionality of the interacting ensemble of donor and acceptor emitters, encoded in the interacting emitters' temporal decay dynamics. While individual emitters decay exponentially, the lifetime decay dynamics of interacting ensemble of emitters follow a stretched exponential decay, revealing a non-integer power $\beta$ in time,
\begin{equation}
    \centering
    I(t)/I_0 = exp(-\gamma_Dt) exp(-\alpha t^\beta)
\end{equation}
where $\gamma_D$ is the spontaneous decay rate and $\alpha$ is the effective interaction volume \cite{forster1949experimentelle, drake1991chemical, klushin2003effects}. The non-integer power, $\beta$,  originates due to DDIs between the emitters and captures the apparent dimensionality sensed by the mutually interacting emitters.
\begin{equation}
\centering
    \beta = \bar{d}/S
\end{equation} $\bar{d}$ is the apparent dimension, and $S = 6$ for electric DDIs \cite{drake1991chemical}. Such relaxation decay dynamics arising due to DDIs are common in other systems such as the kinetic Ising model below the critical temperature, an interacting ensemble of spins \cite{davis2021probing}, in ultra-cold atoms, and ions \cite{neyenhuis2017observation, abanin2019colloquium, ratschbacher2013decoherence, takano1988stretched, kucsko2018critical, hanson2008coherent}. The underlying physics that governs DDIs is universal; here, we focus on DDIs at room temperature, where it is difficult to discern coherent effects. 
\par
The two underlying characteristics that relate to this intriguing non-integer power, $\beta$ in the decay dynamics (and thus the apparent dimensionality) are (i) the distance scaling law associated with DDIs in the vicinity of nanophotonic environment and (ii) the competition between the characteristic DDI length-scale, $R_0$, and the system size, $L_{sys}$. The interplay between these two characteristic lengths determines the spatial extent of the emitters sensed by each donor quantum emitter.  Figure \ref{fig1Concept} conceptually shows the origin of the reduced apparent dimensionality. In homogeneous environments, the DDI potential, $V_{dd}$, scales as $\sim 1/R^3$. The non-integer power, $\beta =  1/2$ $(1/3)$ for the three-dimensional (two-dimensional) spatial distribution of emitters \cite{ boddeti2021long} for time-scales beyond the coherence times of the interacting system (i.e., the emitters do not possess memory of previous interaction events). See supplementary material for more information \cite{SM,hastings1970monte,boddeti2021long, nakamura2004scrutinizing, klushin2003effects, berberan2005mathematical, lee2011programmable, henzie2007multiscale, rackauckas2017differentialequations, novotny2012principles}.
\par 
In contrast to homogeneous environments, a resonant nanophotonic structure modifies the strength, range, and characteristic interaction length scale of DDIs \cite{boddeti2021long, skljarow2022purcell, newman2018observation, cortes2017super, biehs2016long}. Due to this modification of underlying electromagnetic fields, an ensemble of interacting quantum emitters coupled to such resonant nanophotonic structures perceive a modified spatial distribution of emitters. Thus, the spatial extent, strength, and confinement of electromagnetic fields, the hierarchy of distances (and thus the DDI strength) averaging over all possible sites of the interacting emitters is modified. This leads to a modification in the temporal decay dynamics which is reflected in the non-integer exponent, $\beta$, and hence, the apparent dimensionality of the interacting system.
\par
\begin{figure}[!h]
    \centering
    \includegraphics[width = 0.45\textwidth]{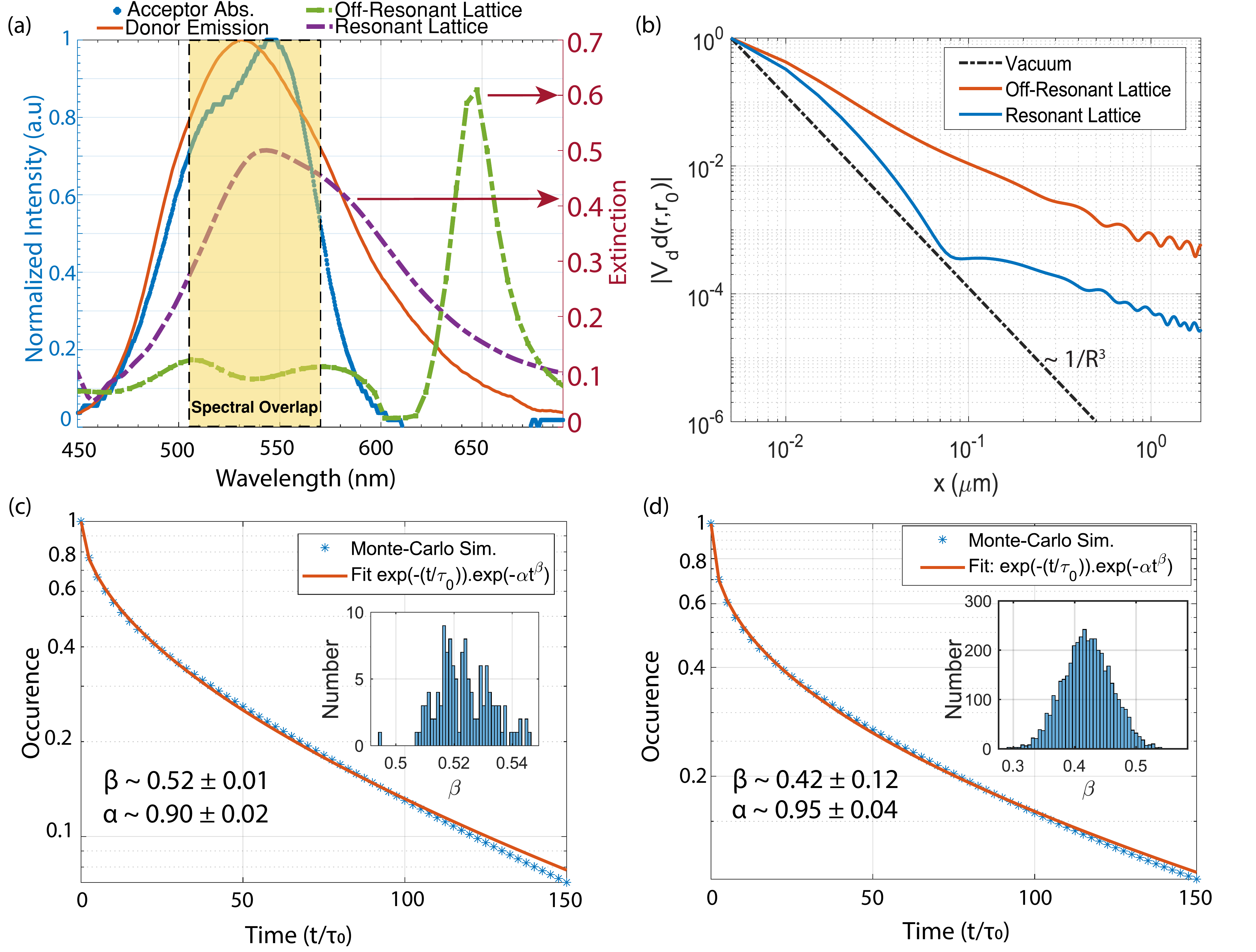}
    \caption{(a) The plot shows the acceptor emitter's absorption spectrum (blue curve), the donor emitter's emission spectrum (orange curve), the extinction spectrum of a resonant plasmonic lattice with lattice constant $\sim$ 300 nm (purple dash curve) and an off-resonant plasmonic lattice with lattice constant $\sim$ 350 nm (green dash-dot curve). The extinction spectrum of the resonant plasmonic lattice spectrally overlaps with the emission-absorption spectrum of the donor and acceptor emitters (yellow highlighted region) (b) The calculated dipole-dipole interaction potential $|V_{dd}|$ for the resonant and off-resonant plasmonic lattice is shown. The resonant plasmonic lattice shows a strikingly modified scaling law. (c) Monte-Carlo simulations depicting the temporal decay dynamics of donor emitters for  $|V_{dd}|^2 = R_0^6/R^6$ scaling and $R_0 \ll L_{sys}$ with $\beta \sim 0.52$. The inset shows the values of $\beta$ for randomized spatial distributions of emitters. (d) Monte-Carlo simulations showing the temporal decay dynamics of donor emitters for  $R_0 \sim L_{sys}$. The reduced dimensionality is evident from the estimated values of $\beta \sim 0.4$.  The inset shows the values of $\beta$ for randomized spatial distributions of emitters.}
    
    \label{fig2}
\end{figure}

\emph{System--} In this study, we consider the interaction of an ensemble of donor ($Alq_3$) and acceptor (R6G) emitters in both resonant and off-resonant nanophotonic structures. The dipole-dipole interactions (DDIs) between the emitters lead to resonance energy transfer. The DDI potential is related to the dyadic Green's function, $V_{dd}(\mathbf{r_A}, \mathbf{r_D}; \omega_D) = -(\omega_D^2/\epsilon_0 c^2)\mathbf{n_A}.\mathbf{\rttensortwo{G}(r_A,r_D;\omega_D)}.\mathbf{n_D}$, where $\mathbf{r_A}$ and $\mathbf{r_D}$ are the positions of the acceptor and donor emitters, respectively, $\mathbf{n_A}$ and $\mathbf{n_D}$ are unit orientation vectors of the acceptor and donor emitters, respectively, $\omega_D$ is the radial frequency of the donor emitter, $\epsilon_0$ is vacuum permittivity, and $c$ is the speed of light \cite{boddeti2021long, cortes2017super}. The interaction strength is proportional to the rate of energy transfer, $\Gamma_{ET}(\mathbf{r_A},\mathbf{r_D};\omega_D) = (2\pi/\hbar^2)|V_{dd}(\mathbf{r_A},\mathbf{r_D};\omega_D)|^2f_D(\omega_D)\sigma_A(\omega_D),$ where $f_D(\omega_D)$ and $\sigma_A(\omega_D)$ are the emission spectra of the donor emitter and absorption cross-section of the acceptor emitter, respectively. Figure \ref{fig2}(a) shows the spectral overlap between the donor emission spectrum (Alq3), the acceptor absorption spectrum (R6G), and the extinction spectrum of both a resonant and an off-resonant plasmonic lattice. The resonant plasmonic lattice modes mediate the DDIs between the donor and acceptor emitters. The resonant plasmonic lattice modifies the scaling, strength, and range of the DDI potential $|V_{dd}|$ as shown in Fig \ref{fig2}(b). The scaling of the DDI potential, $|V_{dd}|$, is significantly modified with distance $R = \mathbf{|r_D - r_A|}$ in a resonant structure, whereas the DDI potential decays rapidly with distance in an off-resonant plasmonic lattice. The resonances of the plasmonic lattice modes can be tuned by altering the lattice constant.
\par
The relaxation dynamics of the interacting ensemble of donor-acceptor emitters are governed by non-linear coupled rate equations (see supplementary material) \cite{SM}. Here the Monte-Carlo simulation method is employed to estimate the temporal decay dynamics of the donor emitters (see supplementary material) \cite{SM}. Figure \ref{fig2}(c) shows the estimated temporal decay dynamics for homogenous environments, i.e., $R_0 \ll L_{sys}$, where non-integer exponent, $\beta \sim 0.52$. This is commensurate to a three-dimensional interacting system and matches well with the predicted theoretical value (see derivation in supplementary material)\cite{SM}. The inset shows the estimated values of $\beta$ for various runs of the Monte-Carlo simulations with different random spatial distributions of the emitters. On the other hand when $R_0 \sim L_{sys}$ as shown in Fig.\ref{fig2}(d), the value of  non-integer exponent, $\beta \sim 0.42(12)$. This is commensurate to an effective dimension of $\bar{d} \sim 2.50(72)$--- a lower than a three-dimensional system. The inset shows the broad distribution in the values of $\beta$ with a standard deviation of  $\sim$ 0.12 for 1024 different iterations of the Monte-Carlo simulation. 
\par
In practice, a resonant plasmonic lattice aide in realizing an apparent lower-dimensional system. The modified scaling of the DDI potential, $|V_{dd}|$ coupled with increased interaction strength, leads to an increase in the characteristic interaction length scale, $R_0$. Under certain conditions when the system size, i.e., the spatial extent of emitters, $L_{sys}$ becomes comparable to the $R_0$ in addition to the scaling law, the interacting system of emitters (in resonant nanophotonic structures) perceive an apparent lower dimension. We explore this effect here to engineer the dimensionality of collective (many-dipole) DDIs. 
  \begin{figure}
    \centering
    \includegraphics[width = 0.42\textwidth]{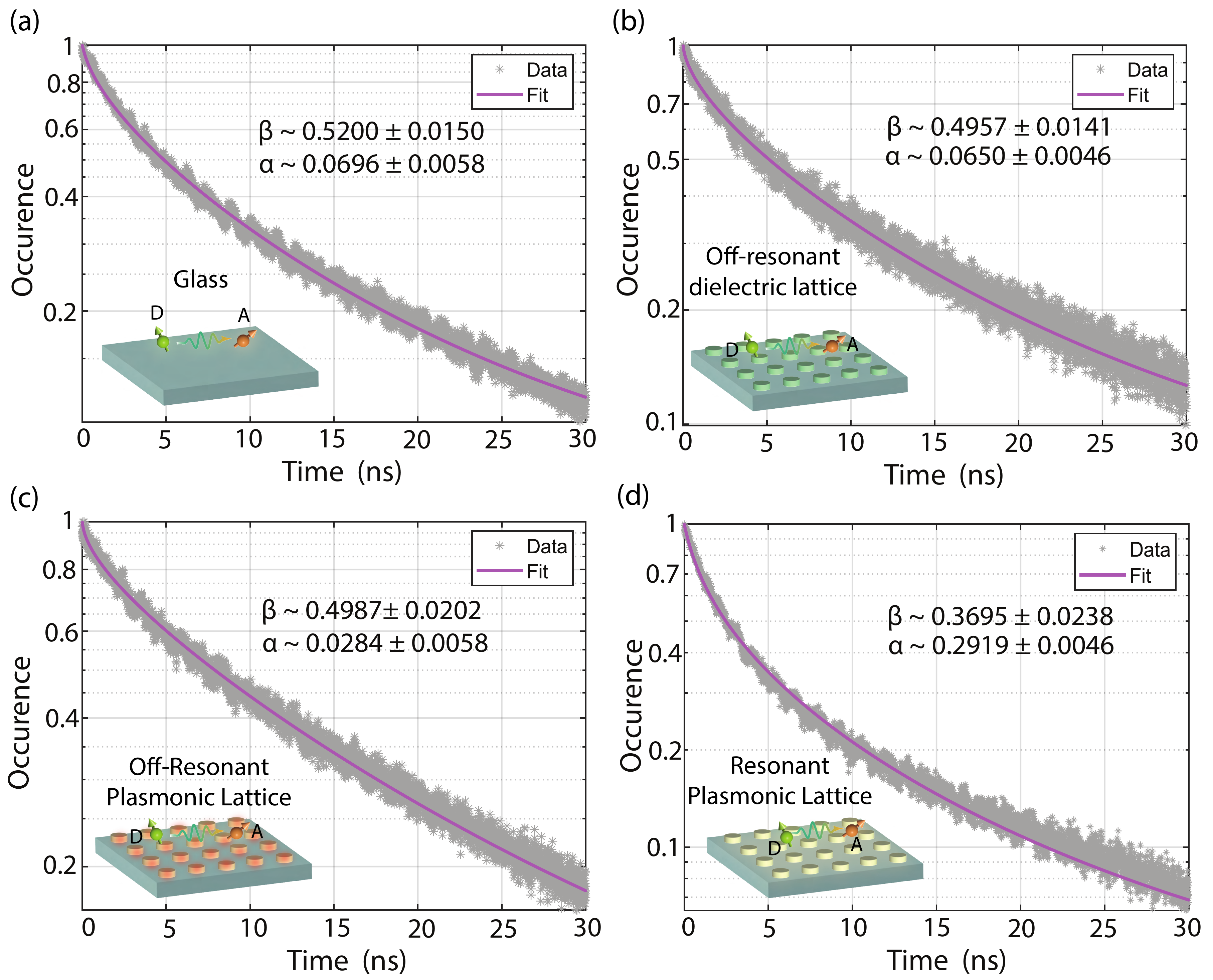}
    \caption{The measured fluorescence lifetime decay  when the interacting
emitters are in different electromagnetic environments (a) glass substrate (i.e, a homogeneous
environment), (b) TiO2 dielectric lattice (i.e. an off-resonant inhomogeneous electromagnetic
environment), and (c) a plasmonic lattice (i.e. a resonant inhomogeneous electromagnetic
environment). The value of $\beta \sim$ 0.5 in both inhomogeneous and off-resonant inhomogeneous
environment. This is commensurate with a 3D system. In contrast, the faster-than-exponential decay dynamics on a resonant silver (Ag) plasmonic lattice reveals an exponent value of $\sim$ 0.37. This is commensurate to an effective lower dimension $\bar{d} \sim 2.20 (12)$. The emitters were embedded in a $\sim$ 1 $\mu m $ thick polymer thin films.}
    \label{fig:LifeTime}
\end{figure}

\par
\emph{Experiment-- }To elucidate this, in the experiment,  we measure the fluorescence lifetime decay trace of the interacting emitters in both resonant and off-resonant nanophotonic structures. The dye molecules $Alq_3$ (0.83 mM) and $R6G$ (0.25 mM) are embedded in PMMA polymer thin films on the aforementioned samples. We use time-correlated single-photon counting technique with a narrow-band filter (520(5) nm) centered at the peak emission of the donor emitter to measure the fluorescence lifetime decay traces (see supplementary material) \cite{SM}.  Figure \ref{fig:LifeTime} shows the measured lifetime decay when the interacting emitters embedded in different nanophotonic structures such as (i) glass substrate, (i.e. a homogeneous environment), Fig.\ref{fig:LifeTime}(a), (ii) a  $TiO_2$ dielectric lattice, Fig.\ref{fig:LifeTime}(b), (iii) an off-resonant plasmonic lattice, Fig.\ref{fig:LifeTime}(c), and (iv) a resonant plasmonic lattice, Fig.\ref{fig:LifeTime}(d).  We observe a striking deviation to the non-integer exponent in time from the typical $\beta =$  0.5 in 3D homogeneous environments to $\beta \sim$ 0.37 (an effective lower dimension $\bar{d} \sim 2.20 (12)$) in a dispersive resonant nanophotonic structure--- a plasmonic lattice.  We note that this value is close to that of a 2D system. This elucidates that the underlying resonant modes supported by the plasmonic lattice indeed modify the apparent dimension perceived by the interacting ensemble of emitters. The $TiO_2$ dielectric lattice has the same geometric features as the resonant plasmonic lattice but supports no resonances. The measurements on the $TiO_2$ lattice help rule out effects due to the underlying geometry of the lattice. On the other hand, the measurements on the off-resonant plasmonic lattice elucidate that the origin of the apparent lower dimension is purely due to the lattice resonance and not from the localized-surface-plasmon-resonance of the constituent metal nanoparticles. 
 \par
 The non-integer exponent in time is estimated by fitting the temporal fluorescence decay trace with a Laplace transform of an underlying probability density function\cite{berberan2005mathematical}, 
\begin{equation}
    \frac{I(t)}{I_0} =  \int^\infty_0 G_\delta(\gamma) e^{-\gamma t}d\gamma \int^\infty_0 H_\beta(\Gamma_{ET}) e^{-\Gamma_{ET} t}d\Gamma_{ET}
    \label{Eq:RDDIExp1}
\end{equation}
In Eq.\ref{Eq:RDDIExp1}, the first term is associated with the spontaneous decay of donor emitters, whilst the second term is associated with resonance energy transfer (DDIs). $G_\delta(\gamma)$ is the probability density function (PDF) associated with the distribution of spontaneous emission decay rates, and $H_\beta(\Gamma_{ET})$ is the PDF for resonant energy transfer rates. For a homogenous environment, with no significant enhancement in the local density of optical states (LDOS), $G_\delta(\gamma) = \delta(\gamma -\gamma_D)$, where $\delta(\gamma -\gamma_D)$ is the delta function, $\gamma_D$ is the decay rate of the individual donor emitter. In contrast, in an inhomogeneous environment, each donor experiences different LDOS and, thus, different spontaneous emission decay rates \cite{akselrod2014probing}. As DDIs in this particular scenario is a weak perturbation, the spontaneous decay rate of the donors, is estimated from the fluorescence decay trace of donors in the absence of the acceptor emitter(see supporting information). The PDF of resonance energy transfer rates, $H_\beta(\Gamma_{ET})$, is estimated by fitting the fluorescence lifetime decay trace with the spontaneous decay rate PDF, $G_\delta(\gamma)$ as a fixed parameter. The underlying probability distributions have a characteristic long-tail behavior and are related to L\'evy stable distributions \cite{nakamura2004scrutinizing}. 
\par
 
\begin{figure}[!h]
    \centering
    \includegraphics[width = 0.4\textwidth]{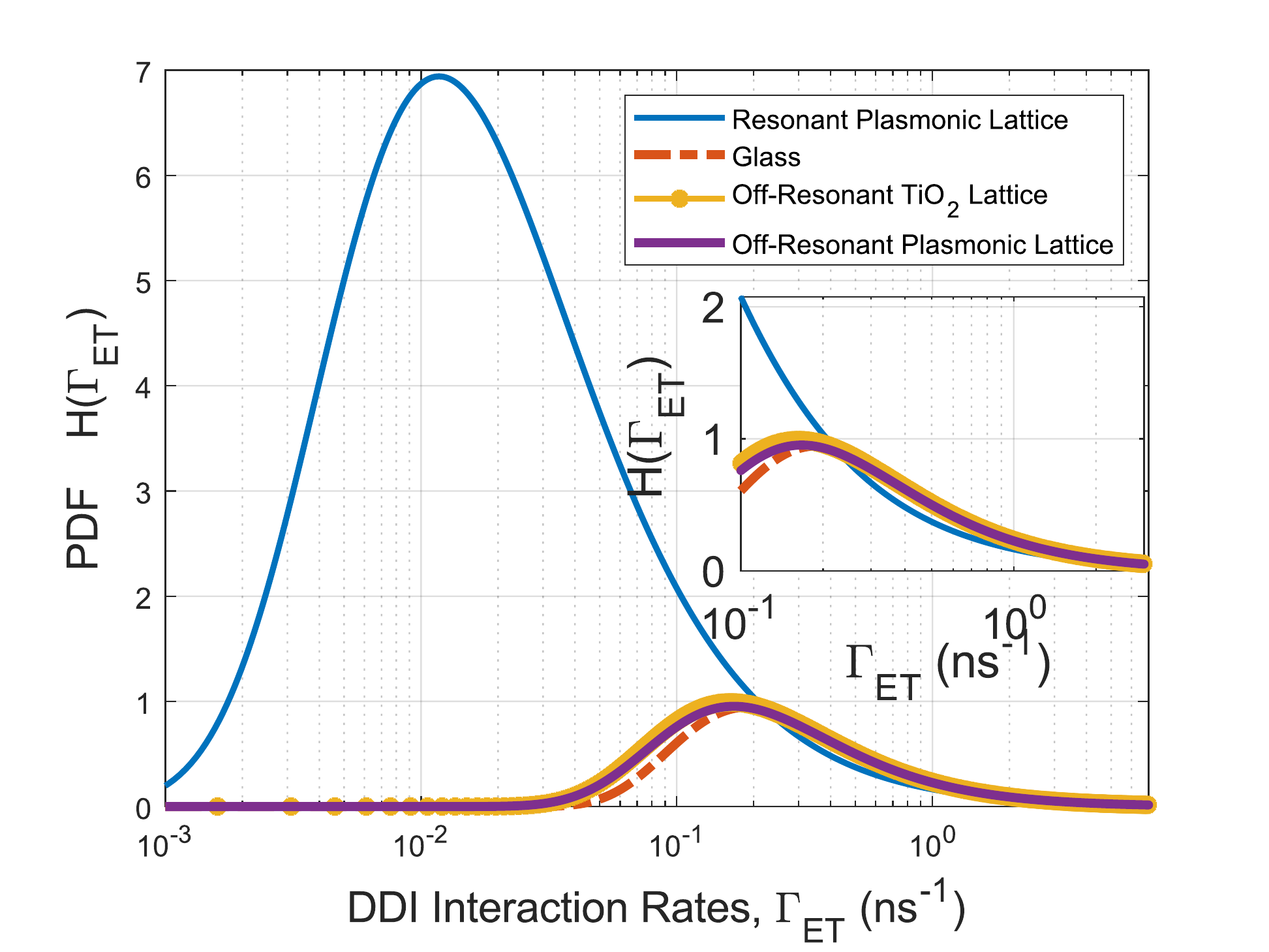}
    \caption{The extracted probability density function (PDF) for the resonance energy transfer rate on various electromagnetic environments (1) Glass, a homogeneous environment (dash-dot red curve), (2) An in-homogeneous environment, $TiO_2$ nanoparticle lattice having the same lattice constant and dimensions as the resonant plasmonic lattice (dot-line yellow curve). (3) An off-resonant plasmonic lattice (purple curve) and (4) A resonant plasmonic lattice (blue curve). The PDF of the energy transfer rates on the resonant plasmonic lattice is not only shifted but also broader. The inset shows the reduced number of events having stringer interaction strength (in the tail)}
    \label{fig:ch3-Ratedistributions}
\end{figure}
Figure \ref{fig:ch3-Ratedistributions} shows the extracted PDF of the resonant energy transfer rate ($\Gamma_{ET}$) distribution. The PDFs obtained in the resonant inhomogeneous environment are observed to differ from those in the homogeneous and off-resonant inhomogeneous environments. This directly indicates that the sensed spatial distribution of emitters is modified.  As the plasmonic lattice supports dispersive delocalized resonant modes that can mediate interactions between the donor and acceptor emitters over larger distances, the underlying PDFs show a broader distribution of rates. Furthermore, the number of interaction events in the tail of the distribution reduces, which indicates a reduction in the total number of larger magnitude DDI interaction strengths, $\Gamma_{ET}$, see inset of Fig. \ref{fig:ch3-Ratedistributions}. The plasmonic lattice shows a higher probability of lower DDI rates. This result may appear counter-intuitive, as the plasmonic lattice enhances the DDI rates overall. The plasmonic lattice enhances the interactions at larger distances, i.e., beyond the F\"{o}rster Resonance Energy Transfer (FRET) radius. The experiments were performed at densities wherein the mean-nearest neighbor separation between the interacting pair of emitters is $\sim 5-10 nm$. At these distances, $\sim 1/r^6$ is more dominant than the scaling arising due to the plasmonic lattice, thus the DDI rates are lower on the plasmonic lattice. The strength of interactions gets enhanced at larger distances as shown in a previous work \cite{boddeti2021long}. See supporting information for more details.
\par
\emph{Conclusion- }In summary, we experimentally demonstrated that the apparent dimensionality of an interacting ensemble of emitters could be modified using a resonant nanophotonic structure. The temporal fluorescence decay dynamics show a non-integer exponent, $\beta$, that relates to the apparent dimensionality of the interacting system. The value of apparent dimensionality on a resonant plasmonic lattice shows a stark contrast value of $\bar{d} \sim 2.20 (12)$, in comparison to $\bar{d} \sim$ 3.0 obtained on glass, an off-resonant $TiO_2$ dielectric lattice, and an off-resonant plasmonic lattice.  Further, we extract the underlying distribution of energy transfer rates for the emitters' interacting ensemble, indicating that the interacting emitters' perceived apparent dimensionality is modified. This arises due to modifying the underlying distribution of energy transfer rates. This work paves the way for engineering interacting systems with apparent lower dimensionality. Though the presented results are semi-classical and discernible coherent effects cannot be observed at room temperatures, they can readily be applied to regimes where quantum effects are more prominent such as in ultra-cold atoms \cite{skljarow2022purcell}, solid-state emitters systems \cite{davis2021probing, dwyer2022probing}, rare-earth ions \cite{ruskuc2022nuclear, uysal2023coherent}, Rydberg excitons in solids \cite{kazimierczuk2014giant}, and quantum-dots systems \cite{tiranov2022coherent}. Such nanophotonic structures can potentially provide an alternative route to realize two-dimensional systems that host new quantum many-body phases, help mitigate decoherence for quantum sensing, memories, and quantum network applications, realize novel, more efficient light-harvesting systems, and potentially improve biological samples imaging.
\par
This work was supported by the U.S. Department of Energy (DOE), Office of Basic Sciences under DE-SC0017717 (A.K.B, A.B, V.S, and Z.J), IUCRC at the US National Science Foundation under Grant No. 2224960, and the Air Force Office of Scientific Research under award number FA9550-23-1-0489 (H.A), the Office of Naval Research (ONR) under ONR N00014-21-1-2289 (Y.W., and T.W.O.) and the National Science Foundation under DMR-2207215 and DMR-1904385 (X.G.J. and T.W.O.). This work made use of the NUFAB and EPIC facilities of Northwestern University’s NUANCE Center, which has received support from the SHyNE Resource (NSF ECCS-2025633), the IIN, and Northwestern’s MRSEC program (NSF DMR-1720139).

\bibliography{apssamp}
\end{document}

\title{Supporting Information:\\ Reducing effective system dimensionality with long-range collective dipole-dipole interactions}

\author{Ashwin K. Boddeti}
\affiliation{Elmore Family School of Electrical and Computer Engineering, Purdue University,
West Lafayette, Indiana 47907, USA}
 \affiliation{Birck Nanotechnology Center, Purdue University,
West Lafayette, Indiana 47907, USA}
\author{Yi Wang}
\affiliation{Graduate Program in Applied Physics,
Northwestern University, Evanston, IL, 60208 USA}
\author{Xitlali G. Juarez}
\affiliation{Department of Materials Science and
Engineering, Northwestern University, Evanston, Illinois 60208,
USA}%
\author{Alexandra Boltasseva}
\affiliation{Elmore Family School of Electrical and Computer Engineering, Purdue University,
West Lafayette, Indiana 47907, USA}
 \affiliation{Birck Nanotechnology Center, Purdue University,
West Lafayette, Indiana 47907, USA}
\author{Teri W. Odom}
\affiliation{Graduate Program in Applied Physics,
Northwestern University, Evanston, IL, 60208 USA},
\affiliation{Department of Materials Science and
Engineering, Northwestern University, Evanston, Illinois 60208,
USA}
\affiliation{Department of
Chemistry, Northwestern University, Evanston, Illinois 60208, USA}
\author{Vladimir Shalaev}
\affiliation{Elmore Family School of Electrical and Computer Engineering, Purdue University,
West Lafayette, Indiana 47907, USA}
 \affiliation{Birck Nanotechnology Center, Purdue University,
West Lafayette, Indiana 47907, USA}
\author{Hadiseh Alaeian}
\affiliation{Elmore Family School of Electrical and Computer Engineering, Purdue University,
West Lafayette, Indiana 47907, USA}
 \affiliation{Birck Nanotechnology Center, Purdue University,
West Lafayette, Indiana 47907, USA}
\affiliation{School of Physics and Astronomy, Purdue University,
West Lafayette, Indiana 47907, USA}
\author{Zubin Jacob} \email{zjacob@purdue.edu} 
\affiliation{Elmore Family School of Electrical and Computer Engineering, Purdue University,
West Lafayette, Indiana 47907, USA}  \affiliation{Birck Nanotechnology Center, Purdue University,
West Lafayette, Indiana 47907, USA}  

\date{\today}
\maketitle

 \section{Dimensionality in homogeneous environments}

In this section, we assume that the characteristic interaction length-scale $R_0$ is finite, and the system size $L_{sys}$ is infinitely large ($R_0 \ll L_{sys}$). In the presence of an acceptor emitter, owing to dipole-dipole interactions (DDIs), the lifetime of the donor emitter is modified. Thus the temporal decay dynamics of the donor emitter in a homogeneous environment are given as,
\begin{equation}
    I_{DA}(t) = I_0 exp(-\gamma_D t) . exp(-\Gamma_{ET}(r_A, r_D,)t),
\end{equation}
where $I_{DA}$ is the fluorescent decay of the donor in the presence of an acceptor emitter, $\gamma_D$ is the spontaneous decay rate of a donor emitter, and $\Gamma_{ET} (r_A, r_D)$ is the resonance energy transfer rate arising due to dipole-dipole interactions between the donor emitter at the position, $r_D$, and acceptor emitter at the position, $r_A$. Here we consider an experimental scenario where the relative position of the ensemble of emitters, $N_D$ donors, and $N_A$ acceptors remain fixed. This is realized in the experiment by embedding the emitters in a polymer matrix (PMMA). Due to the presence of $N_A$ acceptors, the time-resolved fluorescence decay shows larger quenching,
\begin{equation}
        I_{DA}(t) = I_0 exp(-\gamma_D t) . exp(\sum_i^{N_A}-\Gamma_{ET}(r^i_{N_A}, r_D,)t),
\end{equation}
here, we have dropped the summation over all the donor emitter's spontaneous decay, as the spontaneous decay is assumed to be the same for all the donors (due to the homogeneous environment, the LDOS experienced by each donor emitter is the same). For a large number of acceptors, the summation is replaced with a continuum,
\begin{equation} 
  I_{DA}(t) = I_0 exp(-\gamma_D t) . \int \rho(r_A) exp(-t  \Gamma_{ET}(r_A, r_D,)) dr_A
\end{equation}
where $\rho(r_A)$ is the spatial distribution of the acceptor emitters. In the ensemble of static donor and acceptor emitters, the donors perceive a distribution of acceptors at fixed positions. However, each specific donor perceives a different spatial distribution of acceptors with different relative distances. As it is not known apriori if a chosen donor emitter in the excited state interacts with a neighboring acceptor emitter or undergoes spontaneous decay, the measured temporal decay dynamics for a given donor must be averaged with respect to the position of all the acceptor emitters,
\begin{equation}
          I_{DA}(t) = I_0 exp(-\gamma_D t) . \Bigg [ \int dr_A \rho(r_A) exp\Big(-\Gamma_{ET}(r_A, r_D,)t\Big)\Bigg]^{N_A},
          \label{eqn:RDDIMany}
\end{equation}
By averaging over the position of the donor emitters, we get,

\begin{equation}
          I_{DA}(t) = I_0 \int dr_D \rho(r_D) exp(-\gamma_D(r_D) t) .\Bigg [ \int  dr_A\rho(r_A) exp\Big(-\Gamma_{ET}(r_A, r_D,)t\Big)\Bigg]^{N_A},
          \label{eqn:RDDIMany1}
\end{equation}

where $\gamma_D(r_D)$ is the position-dependent spontaneous decay rate of the donor emitters.
\\
In the thermodynamic limit, i.e, for a large number of emitters and a large volume, $N/V = C_a$ Eq. \ref{eqn:RDDIMany1} can be written as,
\begin{widetext}
\begin{equation}
    I_{DA}(t) = I_0\int dr_D \rho(r_D) exp(-\gamma(r_D)t) exp\Bigg[ -C_a \int dr_A (1-exp(-\Gamma_{ET}(r_A, r_D)t)) \Bigg],
    \label{eqn:RDDIManyFinal}
\end{equation}
\end{widetext}
The above Eq. \ref{eqn:RDDIManyFinal}, when solved for various scaling laws, reveals a non-integer exponent in time, t. 
\\
Here we show the result for dipole-dipole interactions, where the power-law scales as $\sim 1/R^3$ for various cases---  1-, 2-, and 3-dimensions. 
\subsection{Dimensionality in thermodynamic limit}
Here we assume a homogeneous environment, i.e. the donors perceive the same LDOS at every position
\begin{itemize}
    \item \textbf{1 Dimension case: } 
    \begin{equation}
        I_{DA}(t)  = \kappa. exp\Bigg(-C_a t^{1/6} \Gamma\Big (\frac{5}{6}\Big )\Bigg), 
    \end{equation}

     \item \textbf{2 Dimension case: } 
    \begin{equation}
        I_{DA}(t)  = \kappa. exp\Bigg(-\pi C_a t^{1/3} \Gamma\Big (\frac{2}{3}\Big )\Bigg), 
        \label{Eq: 2Dbeta}
    \end{equation}
    and,
        \item \textbf{3 Dimension case: } 
    \begin{equation}
        I_{DA}(t)  = \kappa. exp\Bigg(-4\pi^2 C_a t^{1/2} \Bigg), 
        \label{Eq: 3Dbeta}
    \end{equation}
\end{itemize}
where, $\kappa = I_0 \int dr_D \rho(e_D)exp(-\gamma_D(r_D)t)$.
 Here, we have shown that the fractional power in time is dependent on the spatial distribution of the emitters, i.e., distributed in 1D, 2D, or 3D. The fractional exponent in time is 1/6 for 1D, 1/3 for 2D, and 1/2 for 3D spatial distribution of interacting emitters. This simplified view, however, does not adequately describe the resonance energy transfer process and, thus, dipole-dipole interactions in an ensemble of interacting emitters with spatial constraints. The relaxation processes in any system with spatial confinement show a significantly more complex behavior, strongly affected by the details of underlying electromagnetic fields (environment).

\section{Relaxation Dynamics and Monte-Carlo Simulations}
\subsection{Relaxation Dynamics}
In a system of interacting ensemble of donor and acceptor emitters, the relaxation dynamics for each donor and acceptor are governed by the non-linear coupled rate equations as shown below, 
\begin{equation}
    % \frac{dn_i}{dt} = -\gamma^i_D n_i  -  \Omega ^{i,j} n_i \sum_{j=1}^{M}\Gamma^{i,j}_{ET}p_j, 
        \frac{dn_i}{dt} = -\gamma^i_D n_i  -  n_i \sum_{j=1}^{M}\Gamma^{i,j}_{ET}p_j, 
    \label{Eq:rate1}
\end{equation}
\begin{equation}
    % \frac{dp_j}{dt} = \gamma^j_A(1 - p_j)  - \Omega ^{i,j}  p_j \sum_{i=1}^{N}\Gamma^{i,j}_{ET}n_i,
     \frac{dp_j}{dt} = \gamma^j_A(1 - p_j)  -  p_j \sum_{i=1}^{N}\Gamma^{i,j}_{ET}n_i,
    \label{Eq:rate2}
\end{equation}
where $p_j$ is the probability of the $j^{th}$ acceptor emitter being the ground state, $n_i$ is the probability that the $i^{th}$ donor emitter is in the excited state. $\gamma^{i(j)}_{D(A)}$ is the spontaneous decay rate of the $i^{th}$ ($j^{th}$) donor (acceptor) emitter. $\Gamma^{i,j}_{ET}$ is the pairwise resonance energy transfer rate between the $i^{th}$ donor emitter, and the $j^{th}$ acceptor emitter. This pairwise resonance energy transfer rate is distance dependent, i.e., $\Gamma^{i,j}_{ET} \sim R_0^6/R^6_{i,j}$, where  $R_{i,j}$ is the distance between the $i^{th}$ and $j^{th}$ donor and acceptor emitter respectively. The summation of all the acceptors (donors) in Eq. 10 (Eq. 11) accounts for all the possible combinations of pair-wise interactions between the donor and acceptor emitters. 

\par
In the above coupled non-linear rate equations, the $i^{th}$ donor emitter can only be depleted either by undergoing a spontaneous decay process or dipole-dipole interactions (resonance energy transfer process). The associated channel of decay, i.e. spontaneous decay rate or dipole-dipole interaction, is determined based on the value of $\Omega ^{i,j}$ where,  0 or 1, which is in turn based on the probability $\probP (\gamma^i_D, \Gamma^{i,j}_{ET}) = \Gamma^{i,j}_{ET}/ (\Gamma^{i,j}_{ET} + \gamma^i_D)$ given as \cite{hastings1970monte},
\begin{equation}
\centering
  \Omega ^{i,j} =  
\begin{cases}
 1, & \text{if }  \probP (\gamma^i_D, \Gamma^{i,j}_{ET}) \ge x  \\
 0, & \text{otherwise} 
\end{cases} 
\label{Eq:metropolis}
\end{equation}
where $x$ is a random number under a normal distribution. 
\subsection{Monte-Carlo Simulations}
\begin{figure*}[!ht]
    \centering
    \includegraphics[width = 0.8\textwidth]{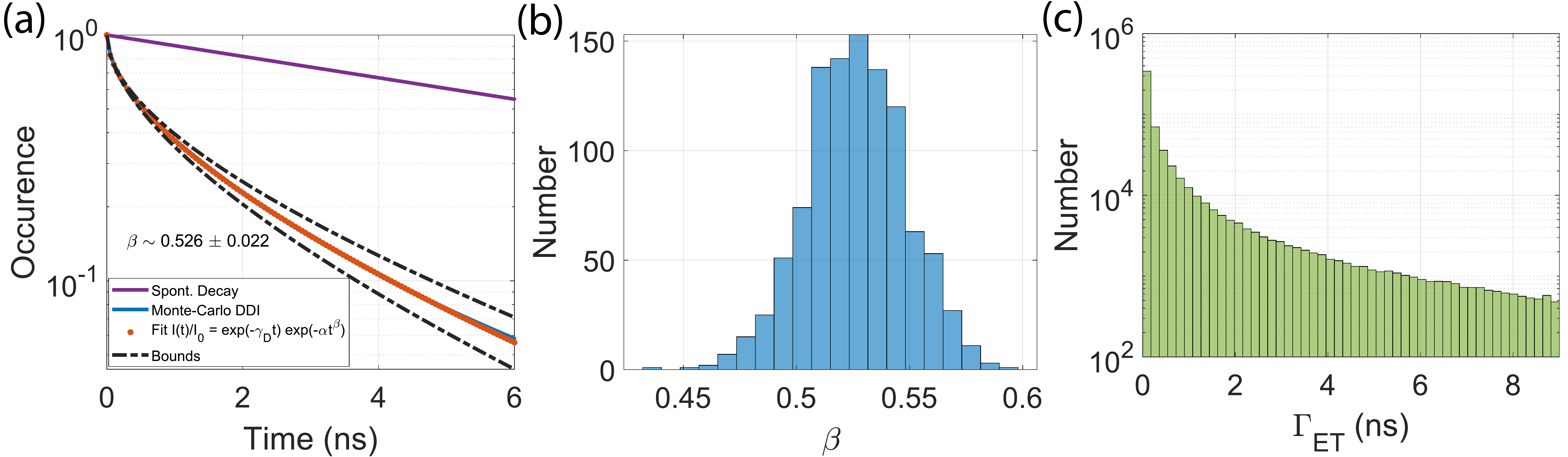}
    \caption{Monte-Carlo simulations showing the fluorescence relaxation decay dynamics of an ensemble of donor emitters undergoing dipole-dipole interactions. (a) Shows the decay dynamics of the donor emitters from the excited state in the absence (spontaneous decay) and presence of acceptor emitters. The donor emitters in the presence of acceptors show a faster than exponential decay (compared to spontaneous decay). This is due to dipole-dipole interactions; an additional decay channel is available to the ensemble of donor emitters. We fit the temporal decay dynamics curve with the stretched exponential (convoluted with a single exponential--spontaneous decay). The bounds are from the standard deviation of 1024 different configurations for which the Monte-Carlo simulations were run Monte-Carlo simulation. (b) The estimated exponent of time, $\beta$, from Monte-Carlo simulations for 1024 configurations. The distribution primarily centers around $\sim$ 0.52. (c) The underlying distribution of energy transfer rates is shown. The distribution of rates is long-tailed.}
    \label{fig:MC-Bulk}
\end{figure*}

The N $\times$ M coupled non-linear rate equations Eq. \ref{Eq:rate1} and Eq. \ref{Eq:rate2} are numerically solved as follows:
\begin{itemize}
    \item N random positions for donor emitter and M random positions for acceptor emitter are chosen in a box of size $L_x$, $L_y$, and $L_z$. The box size considered in the simulations is $50 \times 50 \times 50 nm^3$ for the bulk simulations. 
    \item Pairwise distance between each donor-acceptor pair, $R_{ij}$ is estimated  and the corresponding resonance energy transfer rate, $\Gamma^{i,j}_{ET}$ is calculated. 
    \item We use a Metropolis-Hastings algorithm \cite{hastings1970monte} to determine which among the N donors undergoes a spontaneous decay process as opposed to the dipole-dipole interaction process and vice-versa. To this end, as shown in Eq. \ref{Eq:metropolis} a random number $x$ under a normal distribution is generated and compared to the value of  $\probP (\gamma^i_D, \Gamma^{i,j}_{ET})$. 
    \item The non-linear coupled rate equations Eq. \ref{Eq:rate1} and Eq. \ref{Eq:rate2} are solved using Julia SciML \cite{rackauckas2017differentialequations}. Under certain conditions, the non-linear coupled differential equations were found to be stiff. In such conditions, the QNDF, CVODE-BDF, RADAUIIA2,  and Vern9 algorithms were used to solve the non-linear coupled differential equations in Julia \cite{rackauckas2017differentialequations}.
    \item The Monte-Carlo simulation is run for many different configurations of donor and acceptor emitters. In this case, we have run the simulations for 1024 configurations for the data shown in the main text.
\end{itemize}
In the simulations, we use the density of the donor emitters is, $\rho_D \sim 4.54 \times 10^{24}  molecules/cm^3$ and of acceptors,  $\rho_A  \sim 4.7 \times 10^{24}    molecules/cm^3$. 
Figure \ref{fig:MC-Bulk} (a) shows the relaxation dynamics from Monte-Carlo simulations for an ensemble of donor emitters undergoing dipole-dipole interactions. In the absence of acceptor emitters, the donor emitters decay via a spontaneous decay process (see purple curve), In contrast, in the presence of interaction with randomly distributed acceptor emitters, the donor emitters at shorter time-scales show a faster than exponential decay (see blue curve and the fit). This faster-than-exponential decay arises due to modified decay rates owing to distance-dependent dipole-dipole interactions with acceptor emitters. We extract the effective dimensionality as perceived by the interacting emitters. The value of $\beta \sim 0.53(02)$ which commensurate with the system being 3D as shown from the analytic expression in Eq. \ref{Eq: 3Dbeta}. Figure \ref{fig:MC-Bulk}(b) shows the $\beta$ value estimated for each configuration of the Monte-Carlo simulation runs. We note that the distribution is centered $\sim$ 0.52.  Figure \ref{fig:MC-Bulk}(c) shows the underlying distribution of resonance energy transfer rates. The distributions are long-tailed in nature. This long-tail behavior is associated with rare events contributing to dipole-dipole interactions.

\par 
There are no boundary conditions in our Monte-Carlo simulations. Each configuration and thus the computation box is treated as a separate simulation. As the non-linear coupled equations only have a time-dependence there are no boundary conditions, only initial conditions are needed. In the simulation we consider the donor emitters to be in the excited state, i.e. $n_i(0) = 1.0$ and the acceptor emitters are in the ground state, i.e. $p_j(0) = 1.0$. Note that $p_j$ is defined as the probability of the acceptor being in the ground state. Furthermore, we have not found any dependence on the box size on the values of $\beta$. 
\par
\begin{figure*}[!ht]
    \centering
    \includegraphics[width = 0.7\textwidth]{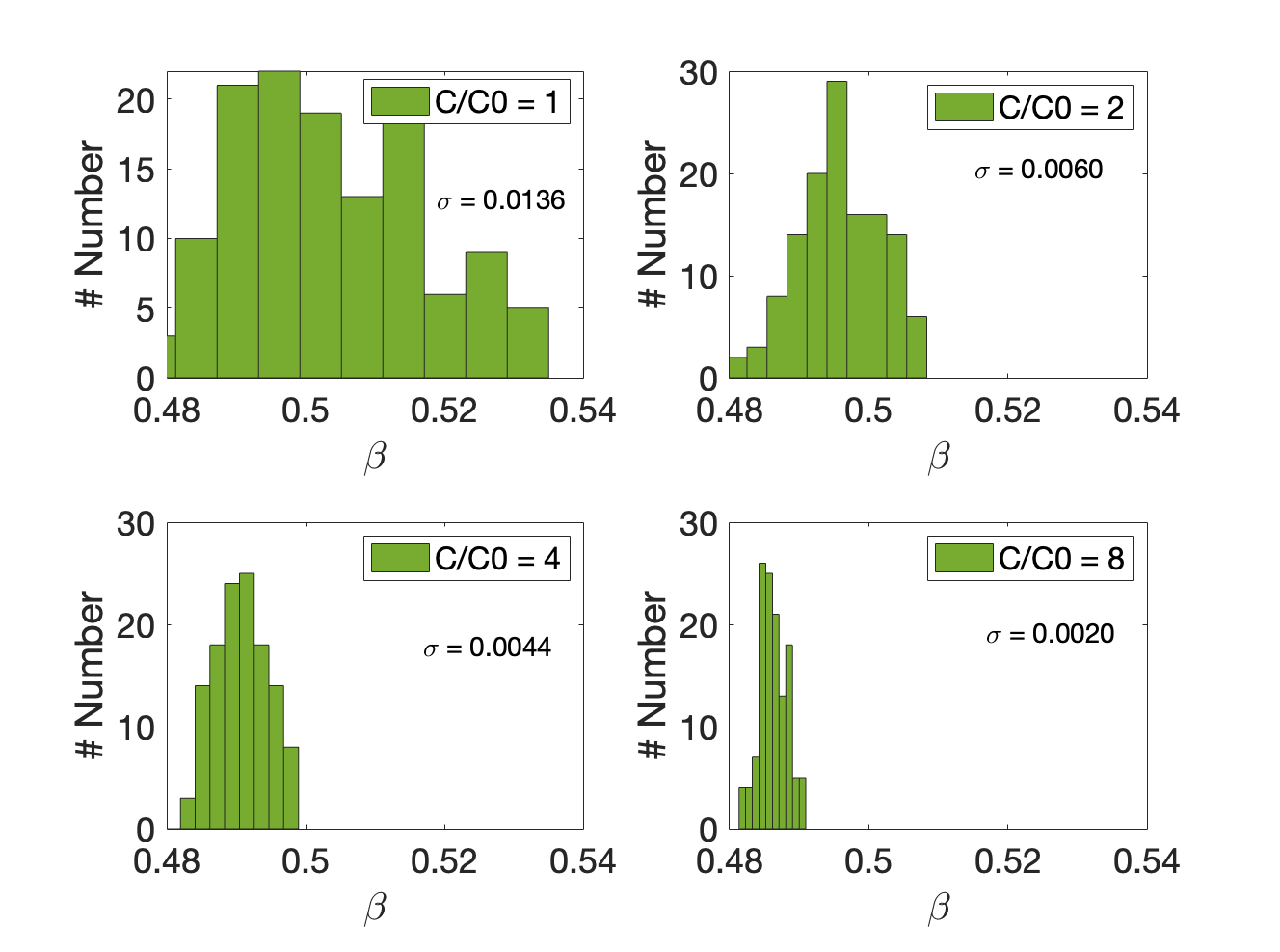}
    \caption{Bulk behavior estimated exponent of time, $\beta$, from Monte-Carlo simulations for different densities. Note the reduction in the width of the distributions. $\sigma$  is the standard deviation in the values of $\beta$ for 128 different configurations of the Monte-Carlo simulations.}
    \label{fig:MC-Bulk-beta}
\end{figure*}
\begin{figure*}[!ht]
    \centering
    \includegraphics[width = 0.7\textwidth]{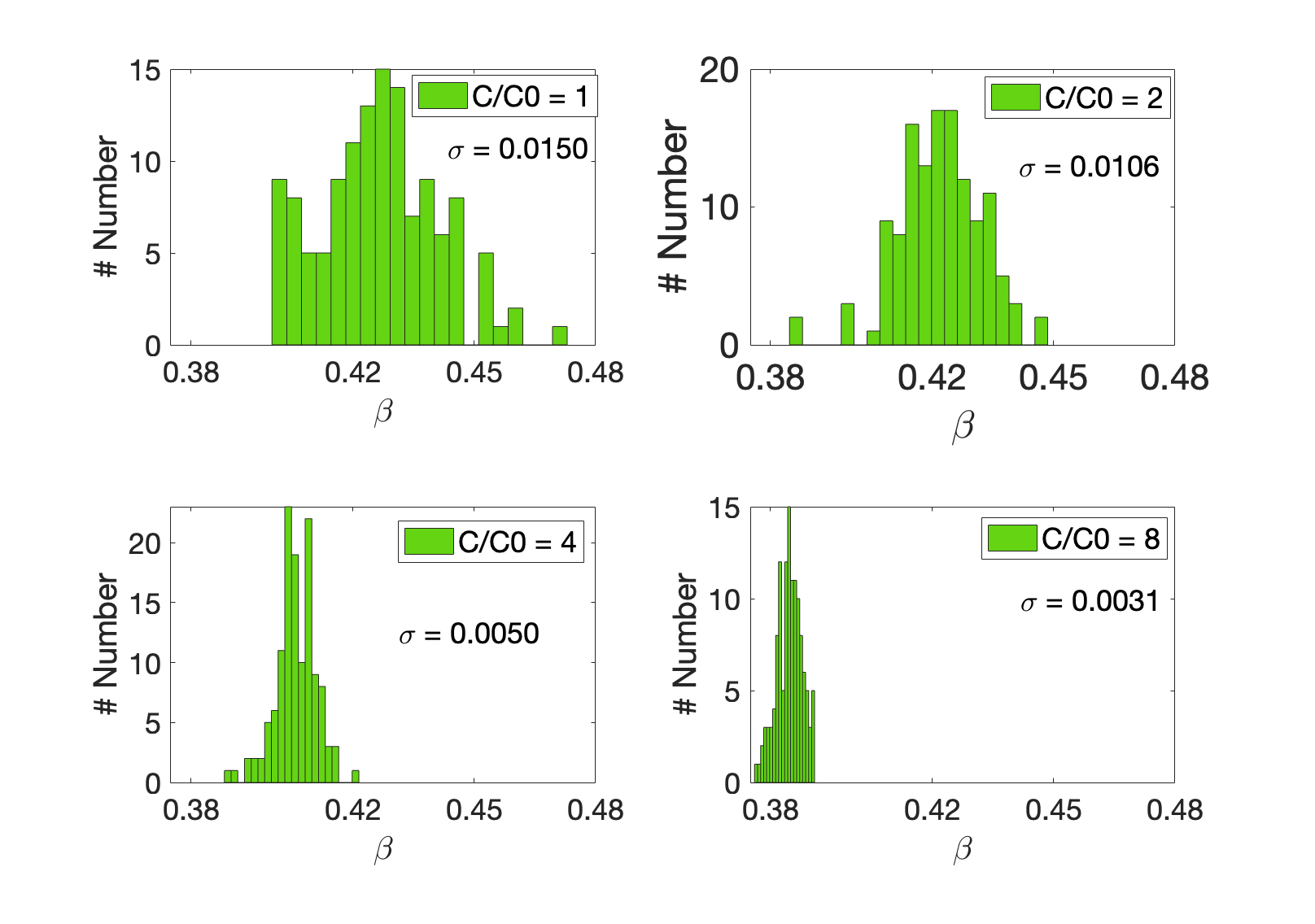}
    \caption{Reduced dimensionality behavior estimated exponent of time, $\beta$, from Monte-Carlo simulations for different densities. Note the reduction in the width of the distributions. $\sigma$  is the standard deviation in the values of $\beta$ for 128 different configurations of the Monte-Carlo simulations.}
    \label{fig:MC-LowDim-beta}
\end{figure*}

The simulations were performed at lower densities than the densities used in the experiment owing to the fact that the number of emitters was very large to simulate. The distribution in the value of $\beta$ occurs due to the Monte-Carlo simulation densities not being in the thermodynamic limit approximation. The distribution in the values of apparent dimensionality reduces as we increase the density of the donor and acceptors in the simulations. This is attributed to the fact that as one increases the densities, the thermodynamic limit is being approached which leads to a reduction in the variation in $\beta$ between various configurations as shown in Fig. \ref{fig:MC-Bulk-beta} and Fig. \ref{fig:MC-LowDim-beta} for the bulk and reduced dimensionality case respectively. 

\begin{figure}[!h]
    \centering
    \includegraphics[width = 0.5\textwidth]{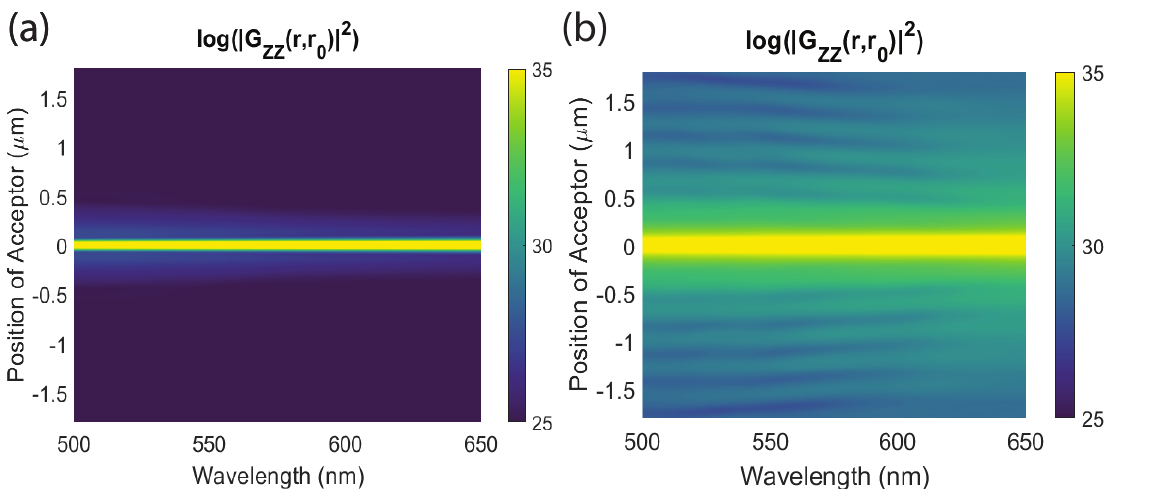}
    \caption{(a) Computed dyadic Green's function across different wavelengths for an off-resonant lattice (b) Computed dyadic Green's function across different wavelengths for a resonant lattice. }
    \label{nearfield}
\end{figure}

\section{FDTD Numerical Simulations}
\subsection{Computing the dyadic Green's function}
The dyadic Green's function $\mathbf{\rttensortwo{G}}$, is defined by the electric field at a position $\mathbf{r_A}$ generated by a point source at position $\mathbf{r_D}$ with dipole moment $\mathbf{\mu}$ \cite{novotny2012principles},
\begin{equation}
    \centering 
    \mathbf{E}(\mathbf{r_A}) = \frac{\omega^2}{\epsilon_0\epsilon_rc^2}\mathbf{\rttensortwo{G}}(\mathbf{r_A},\mathbf{r_D}).\mathbf{\mu}
\end{equation}
Each component of $\mathbf{\rttensortwo{G}}$ is calculated using the corresponding orientation of the dipole and electric field component as
\begin{equation}
    \centering
    G_{xx} = \frac{\epsilon_0\epsilon_rc^
2}{\mu\omega^2}E_{xx}
\end{equation}
The off-diagonal elements are found by taking the corresponding electric field components for a given orientation of the electric dipole. 
The dipole-dipole interaction potential is calculated as
\begin{equation}
\centering
V_{dd} = \frac{-\omega_D^2}{\epsilon_0c^2}\mathbf{n_A}.\mathbf{\rttensortwo{G}(r_A,r_D;\omega_D)}.\mathbf{n_D}
\end{equation}
where $\omega_D$ is the emission frequency of the donor emitter,  c is the speed of light, $\epsilon_0$ is the permittivity, $\mathbf{n_A}$, and, $\mathbf{n_D}$ are the unit vectors corresponding to the donor dipole and acceptor dipole orientations.
Using FDTD simulations, the electric field due to a point dipole placed on top of the structure is estimated. Subsequently, the components of the dyadic Green's function are estimated by running the simulation with a test dipole oriented along x-,y-, and z-directions. A 2$\mu m$ $\times$ 2$\mu m$ structure is simulated with perfectly-matched-layer (PML) boundaries. A  uniform index environment is necessary to excite the collective lattice mode, and this is incorporated into the FDTD simulations by changing the background index. Figure \ref{nearfield} shows the near-field intensity for a resonant and off-resonant plasmonic lattice.
\begin{figure}[!h]
    \centering
    \includegraphics[width = 0.5\textwidth]{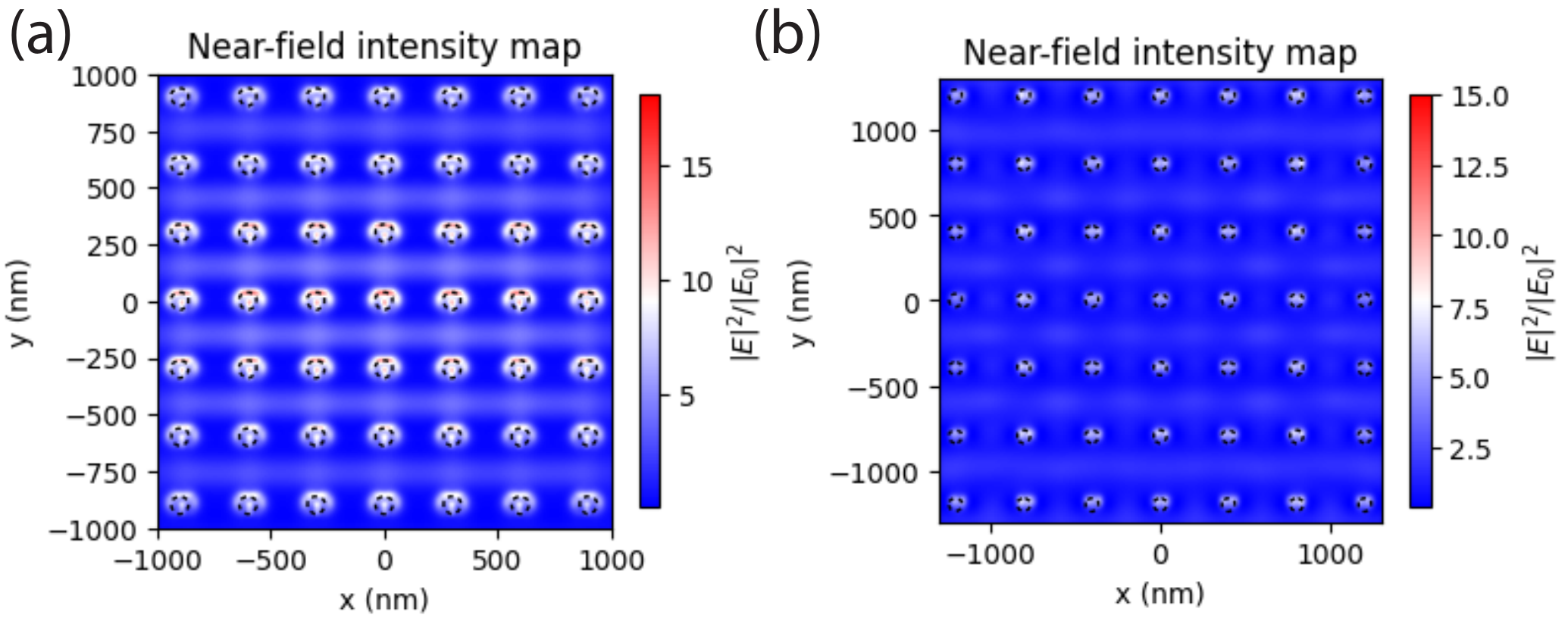}
    \caption{(a) Near-field intensity map for a resonant nanophotonic plasmonic lattice (b) Near-field intensity map for an off-resonant nanophotonic plasmonic lattice }
    \label{nearfield}
\end{figure}
\par
\subsection{Out of plane field confinement}
The dipole-dipole interactions between the ensemble of donors and acceptors are mediated by the surface-lattice resonance (SLR). In the experiment, the emitters are in a thickness of $\sim 1 \mu m$ above the metal nanoparticle lattice. The out-of-plane confinement of the SLR mode determines the effect dimensionality across the $\sim 1 \mu m$ thick emitter distribution. The SLR mode extends to over a $\sim 1 \mu m$ distance. This ensures that all the emitters experience the lattice mode and the interactions are mediated by the SLR mode. 
\begin{figure}[!h]
    \centering
    \includegraphics[width = 0.6\textwidth]{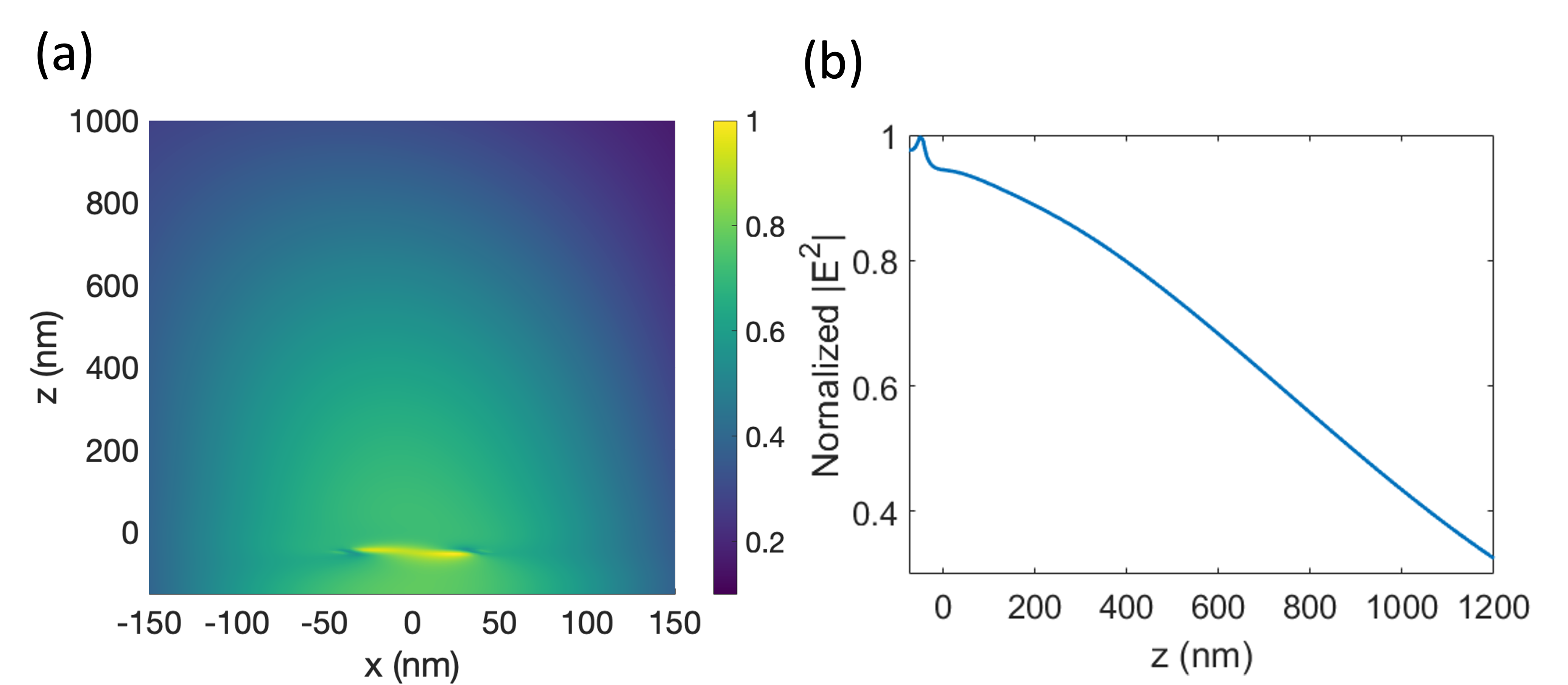}
    \caption{ (a) Normalized out-of-plane confinement of the SLR mode mediating dipole-dipole interactions. The mode extends to $\sim 1 \mu m$ distance. (b) Averaged (along x-direction) normalized field.}
    \label{out-of-plane}
\end{figure}
\section{Experimental Setup}
\begin{figure}
    \centering
    \includegraphics[width = 0.5\textwidth]{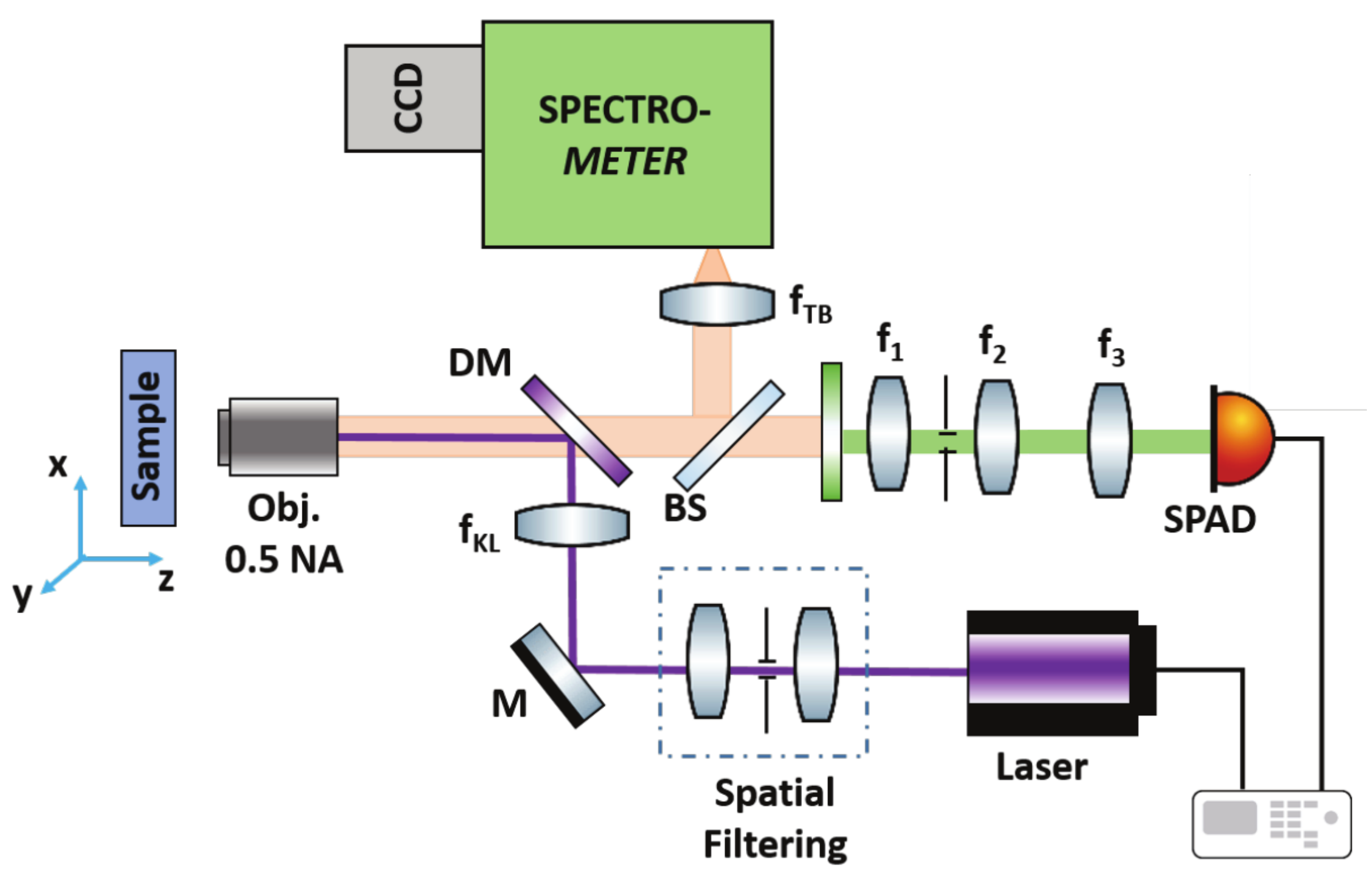}
    \caption{The schematic shows the experimental setup used for lifetime and spectral measurements. DM: Dichroic mirror, M: Mirror, $f_{KL}$: K$\ddot{o}$hler illumination lens, $f_{TB}$: tube lens, $f_1, f_2$ and $f_3$ are achromatic lenses.}
    \label{fig1}
\end{figure}

The spectrum and lifetimes are measured using the experimental setup shown in Fig.\ref{fig1}. The dye molecules are confocally excited using a 405 nm pulsed laser with $\sim 23$ ps pulse width (Alphalas Picopower-LD-405). A K$\ddot{o}$hler illumination lens ($f_{KL}$) is used behind the objective (NA = 0.5) to ensure uniform illumination and prevent photo-bleaching of the donor ($Alq_3$) dye molecules. The collected fluorescence is used for measuring the spectrum, and the decay traces. A narrowband filter (520 (5) nm) is used to collect photons at the peak emission wavelength of donors. The decay trace is recorded using the time-correlated single photon counting (TCSPC) technique.

\begin{figure}[!h]
    \centering
    \includegraphics[scale = 0.3]{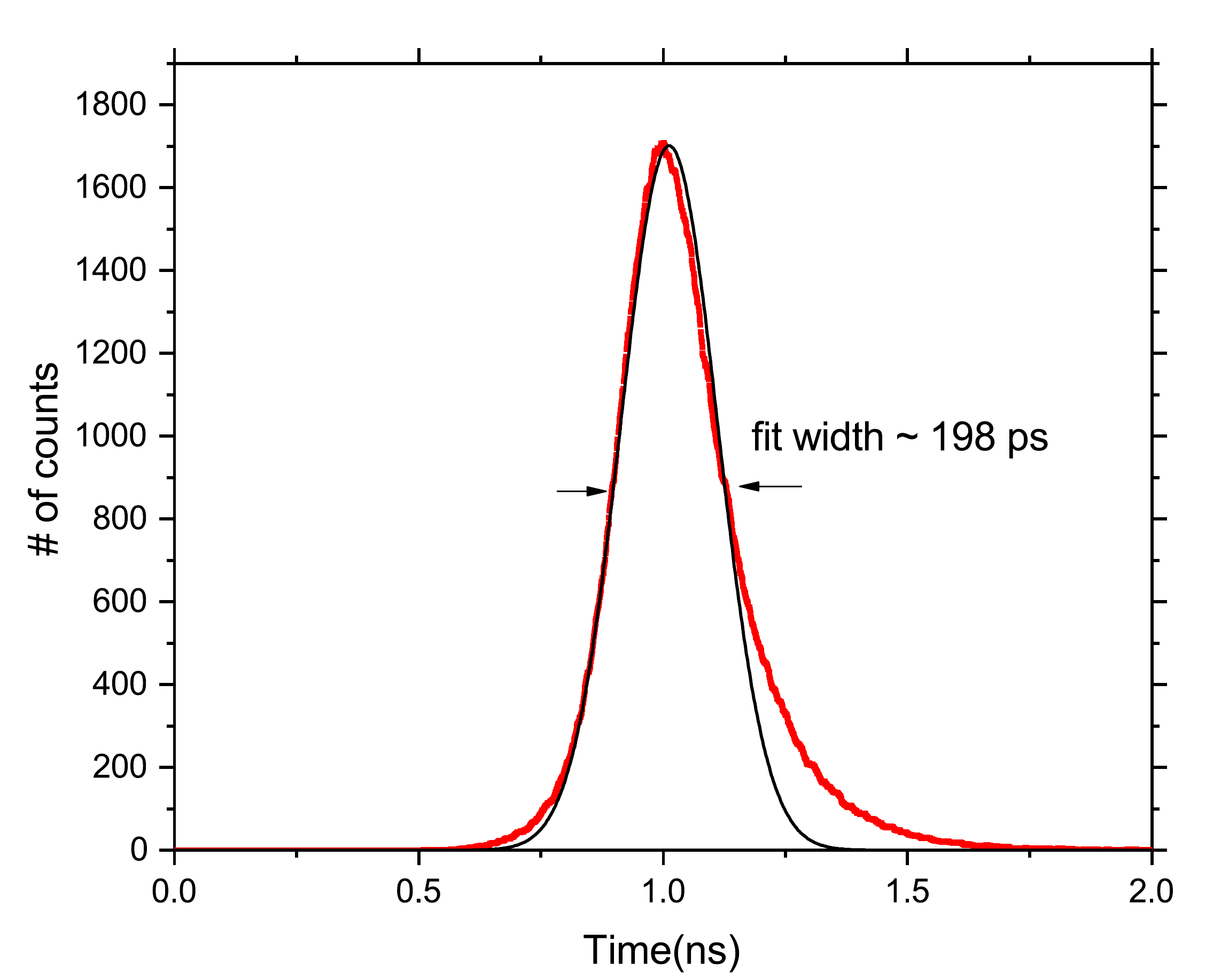}
    \caption{The instrument response function of the TCSPC setup.}
    \label{figtcspc}
\end{figure}
A single photon avalanche diode (MPD PDM series) is used in conjunction with a photon arrival counter board (PicoQuant HydraHarp 400) to detect and time tag the arrival time of the photons after laser excitation pulse (pulse-width $\sim$23 ps). The fluorescence lifetime decay trace is found by binning the detected single photons based on their arrival time after the excitation pulse. Figure \ref{figtcspc} shows the instrument response function (IRF) of the TCSPC set-up used in the lifetime measurements. The IRF is recorded by detecting the reflection of the laser excitation pulses from a glass substrate. A Gaussian function fit estimates the IRF to be $\sim$198 ps FWHM. The measured IRF
results from the convolution of the response functions of the SPAD, specified to be $\sim$35 ps
FWHM by the manufacturer of the acquisition system (electronics), and of the laser pulse duration.

\section{Sample Fabrication}
\begin{figure}[!h]
    \centering
    \includegraphics[width = 0.35\textwidth]{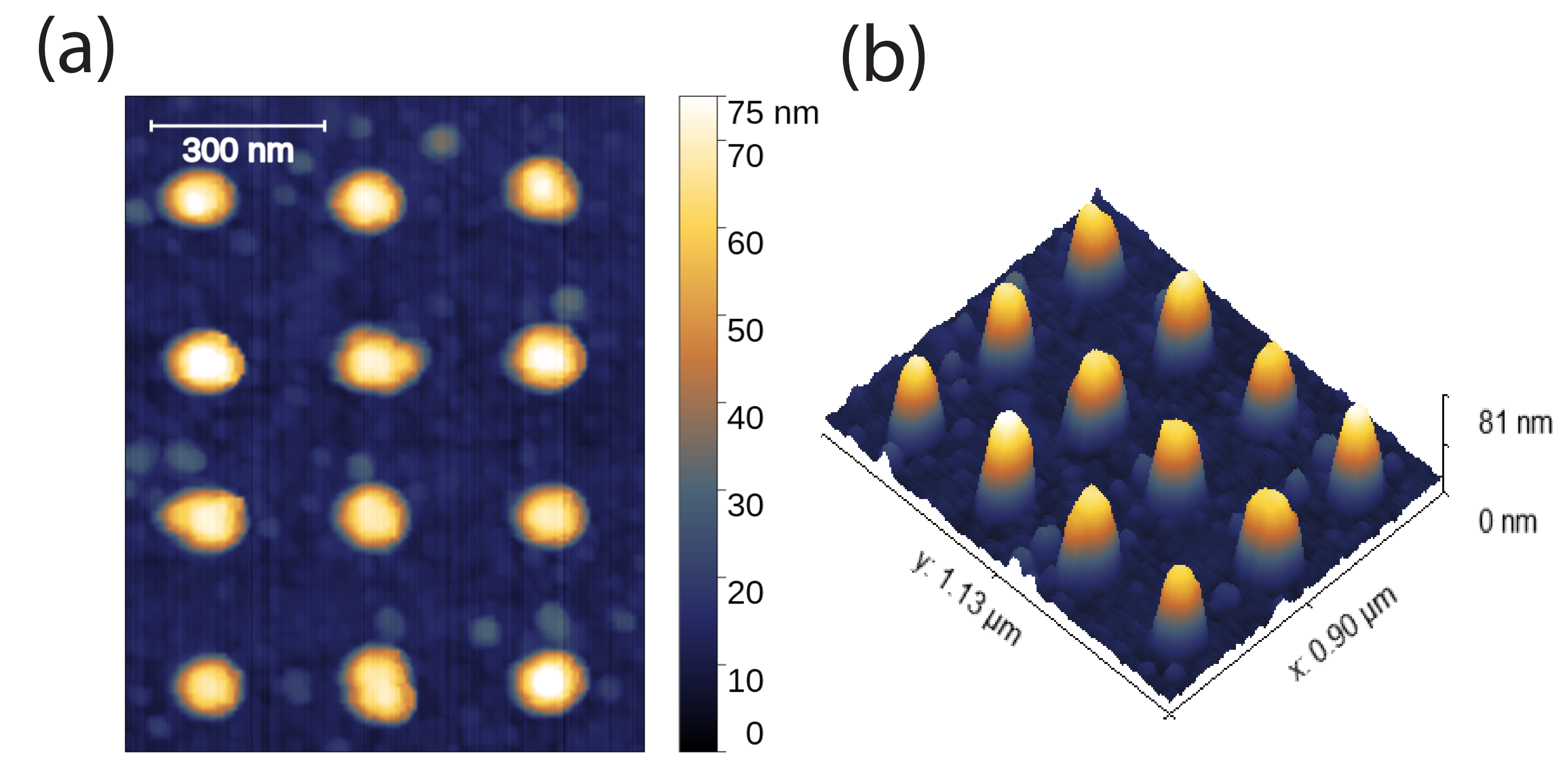}
    \caption{Scanning Atomic Force Microscope image of the sample. }
    \label{fig31}
\end{figure}
Silver (Ag) plasmonic nanoparticle lattices were fabricated through a soft nanofabrication process \cite{henzie2007multiscale}. First,
periodic photoresist posts on Si wafers were generated by solvent-assisted nanoscale embossing (SANE) \cite{lee2011programmable}.  Next, an 8-nm Cr layer was deposited on the Si substrate, followed by lift-off of the photoresist posts and reactive ion etching to create Si pits ( $\sim$ 300 nm) beneath the circular holes in the Cr layer. Au nanohole films that functioned as deposition masks were produced by depositing 130 nm of Au on the substrates, released from the Si wafer by wet etching of the Cr layer, and then transferred and dried on glass substrates. Ag (80 nm) was deposited via electron-beam evaporation through the Au hole mask and then the film was removed with adhesive tape to
produce Ag nanoparticle lattices (area, ~1 $cm^2$ ) on glass substrates. A 5 nm $Al_2O_3$ layer was deposited on the Ag nanoparticle lattices to prevent oxidation and sulfidation of the Ag nanoparticles. Figure \ref{fig31} shows the scanning Atomic Force Microscope image of the fabricated structures. The $TiO_2$ samples were fabricated using the same technique except that no $Al_2O_3$ layer was deposited for protection.

\section{Probability Density Function of Resonance Energy Transfer Estimation}
The luminescence decay can be written as,
\begin{equation}
\centering
I(t) = \int_0^\infty H(\gamma) exp(-\gamma t)d\gamma
\end{equation}
with $I(0) = 1$. Here $H(\gamma)$ is the inverse Laplace transform of $I(t)$ and $\gamma > 0$. $H(\gamma)$ is the normalized probability density function (PDF) of the underlying distribution of rates, as $I(0) = 1$, $\int_0^\infty H(\gamma) d\gamma = 1$. Resonance energy transfer between donor and acceptor molecules exhibits a stretched exponential decay, 
\begin{widetext}
    \begin{equation}
    \centering 
    I(t)  \equiv exp(-\gamma_D t).exp(-\alpha t^\beta) \equiv exp(-\gamma_D t)\int_0^\infty exp(-H_\beta(\gamma)t) d\gamma
    \label{eq:GlassDDI}
\end{equation}   
\end{widetext}
where the first single exponential term represents the intrinsic decay (spontaneous emission) of the donor emitter and the second term corresponds to the distribution of resonance energy transfer rates that occur to dipole-dipole interactions between the donor and acceptor emitters. The determination of $H(\gamma)$ for a given normalized temporal decay trace $I(t)$ requires the computation of the respective Laplace transform. The underlying PDF, $H_\beta(\gamma)$ is given as \cite{berberan2005mathematical},
\begin{equation}
\centering
H_\beta(\gamma) = \tau_0 \frac{B}{(\gamma\tau_0)^{(1-\beta/2)/(1-\beta)}}exp\Bigg[-\frac{(1-\beta)\beta^{\beta/(1-\beta)}}{(\gamma\tau_0)^{\beta/(1-\beta)}}\Bigg]f(\gamma)
\end{equation}
where the function, $f(\gamma)$, is,
    \[
    f(\gamma) = 
    \begin{cases}
        1/[1+C(\gamma\tau_0)^\delta], & \delta = \beta(0.5 - \beta)/(1-\beta), \\ \beta \le 0.5 
        \\
        1 + C(\gamma\tau_0)^\delta, & \delta = \beta(\beta - 0.5)/(1-\beta), \\ \beta > 0.5
        
    \end{cases}
    \]
The parameters $B$ and, $C$ are functions of $\beta$ \cite{berberan2005mathematical}. For the intermediate values, we use a polynomial interpolation function. The function $H_\beta(\gamma)$ is a L\'evy stable distribution \cite{nakamura2004scrutinizing}. While Eq.\ref{eq:GlassDDI} is valid for homogenous environments where every donor emitter in the ensemble experiences the same local density of optical states (LDOS). However, in inhomogeneous environments, each donor emitter experiences different LDOS. Thus, the temporal decay trace is modified. We fit the normalized temporal decay trace with the Laplace transform of the above-defined  PDF function. In the numerics, the limits on the maximum value of rate are chosen such that $\int_0^{\gamma_{max}} H(\gamma)d\gamma = 1 $.
\par
\begin{figure*}[!ht]
    \centering
       \includegraphics[width = 1\textwidth]{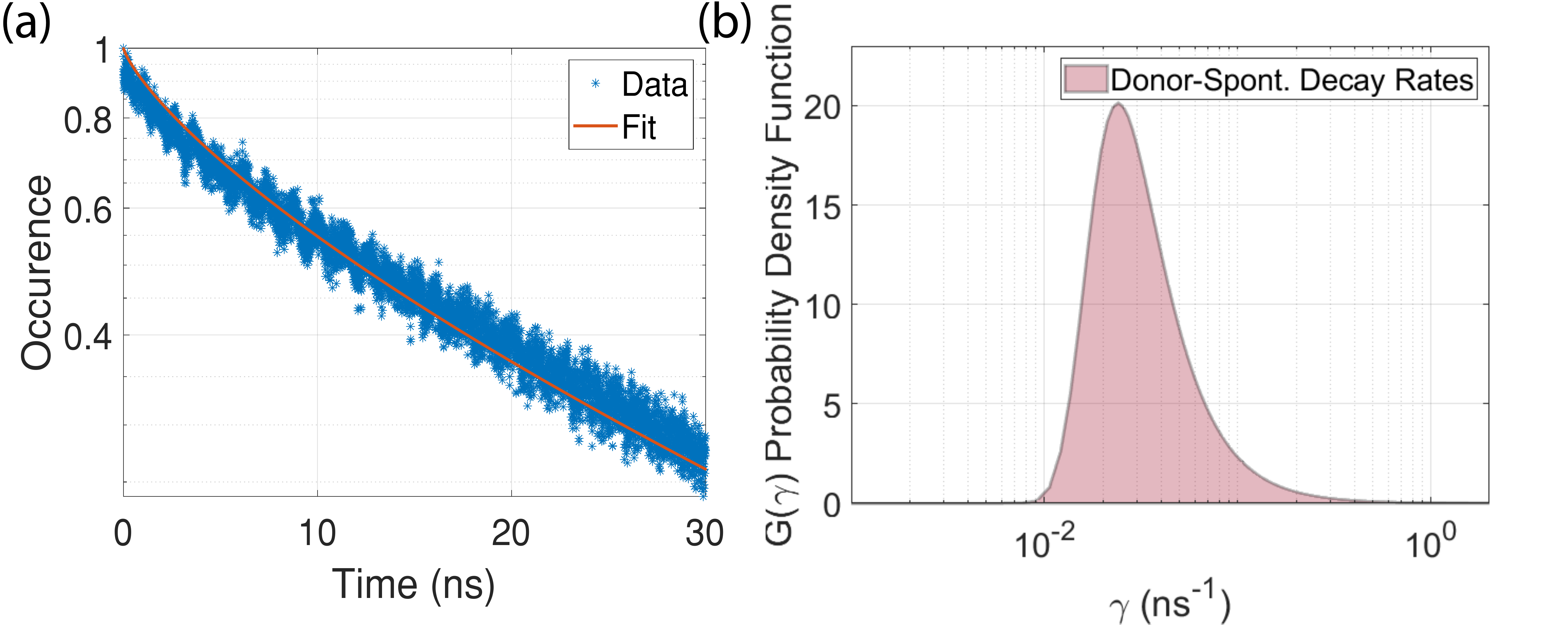}
        \caption{Spontaneous decay of donor emitters, in absence of acceptors in an in-homogeneous resonant environment--the plasmonic lattice. (a) Show the measured fluorescence lifetime decay measurements of the donor emitters on the plasmonic lattice. As the LDOS experienced by each donor emitter is different, the decay dynamics show a stretched exponential behavior. (b) Extracted probability density function of spontaneous emission rate for donors on the plasmonic lattice. A Laplace transform is fit to fluorescence lifetime decay trace to estimate the probability density function $G_\delta(\gamma_D)$}
    \label{fig:SpontDecayRates}
\end{figure*}

 As the interaction is treated as a perturbation, one can find $G_\delta(\gamma_D)$ from fluorescence time decay measurements of donor emitters in the absence of acceptors. The estimated $G_\delta(\gamma_D)$ from non-interacting temporal decay trace are used to extract the resonance energy transfer rates. Figure \ref{fig:SpontDecayRates} shows the decay trace of the donor emitters (no acceptor emitters) alone on the plasmonic lattice. The decay dynamics show a stretched exponential behavior. The underlying PDF for spontaneous decay rates is extracted by using a Laplace transform given as,
 \begin{equation}
     \centering
     \frac{I_D(t)}{I_0} = \int_0^\infty G_\delta(\gamma_D)exp(-\gamma_D t)d\gamma_D
 \end{equation}

In homogeneous environments, we estimate $H_\beta{\Gamma_{ET}}$ by fixing $G_\delta(\gamma_D)$.
\par
\subsection{Lower DDI rates on plasmonic lattice}
The plasmonic lattice enables in enhancement of the range of interactions and the strength of the interactions at larger mean-nearest neighbor distances between the donor and acceptor emitters. In this work, we have performed measurements at densities wherein the mean nearest neighbor distance between the donor and acceptors is $\sim 3nm-5nm$. At these distances the $\sim 1/r^6$ decay of the energy transfer rate (and thus the $\sim 1/r^3$ scaling-law) is more dominant than the long-range interactions scaling. Thus, we notice that the interaction strength on glass is larger than the interaction strength on the plasmonic lattice. However, since the plasmonic lattice supports long-range interactions, the donor emitters can perceive acceptor emitters that are at a much larger distance from them, leading to a broader and an effective lower-strength interaction on the plasmonic lattice. We have studied this effect in detail in a previous work \cite{boddeti2021long}.
\par 
The concentration of the donor molecules is 0.86mM and that of the acceptor molecules is 0.25mM. In this work we have focussed on concentrations C/C0 = 1, here C0 is the acceptor concentration, ${C_0}$ =  0.25mM  (as shown in the figure below from the paper stated above) to study the reduced dimensionality effect. At the highest concentration, we can see that the interaction strength of dipole-dipole interactions on the plasmonic lattice is indeed lower than that noticed on glass. Thus, in this context, though it may be counterintuitive the results are consistent. This counterintuitive result occurs because, at a short distance less than $\sim 5 nm$, the $\sim 1/r^3$ scaling-law of the dipole-dipole interactions ($\sim 1/r^6$  scaling for DDI rates) is dominant than the effective scaling that we notice on the plasmonic lattice. Thus, the interactions on the plasmonic lattice at these densities (donor $\sim 4.54 \times 10^{28} molecules/m^3$ and acceptor $\sim 4.7 \times 10^{28} molecules/m^3$) are lower than that on glass.

% \section{Acceptor Decay Dynamics}
% The acceptor decay dynamics is shown in Fig. \ref{fig:AcceptorSpontDecayRates}

% \begin{figure*}[!ht]
%     \centering
%        \includegraphics[width = 0.6\textwidth]{Hybrid-560nm.png}
%         \caption{Decay dynamics of acceptor emitter at acceptor's peak emission (560 $\pm$ 5 nm) on glass and plasmonic lattice due to interactions with the donor emitters.}
%     \label{fig:AcceptorSpontDecayRates}
% \end{figure*}
%\input{SupplementaryMat.bbl}

%\bibliography{dwrreview}
\bibliography{apssamp}